# Quantum Phase Dynamics in an *LC* shunted Josephson Junction

Ch. Kaiser[1], T. Bauch[2], F. Lombardi[2], M. Siegel[1]

[1]Institut für Mikro- und Nanoelektronische Systeme, Karlsruher Institut für Technologie (KIT), Hertzstraβe 16, D-76187 Karlsruhe, Germany
[2]Department of Microtechnology and Nanoscience, Chalmers University of Technology, SE-412 96, Göteborg, Sweden

We have studied both theoretically and experimentally how an *LC* series circuit connected in parallel to a Josephson junction influences the Josephson dynamics. The presence of the shell circuit introduces two energy scales, which in specific cases can strongly differ from the plasma frequency of the isolated junction. Josephson junctions were manufactured using Nb/Al-AlO$_x$/Nb fabrication technology with various on-chip *LC* shunt circuits. Spectroscopic measurements in the quantum limit show an excellent agreement with theory taking into account the shunt inductance and capacitance in the Resistively and Capacitively Shunted Junction model. The results clearly show that the dynamics of the system are two-dimensional, resulting in two resonant modes of the system. These findings have important implications for the design and operation of Josephson junctions based quantum bits.

I. INTRODUCTION

Josephson junctions (JJs) are considered to be the main building blocks for a variety of exciting applications, ranging from classical electronics [1] to quantum computing [2]. The phase dynamics of JJs have been well understood for low critical temperature superconductors (LTS). Here, the dynamics of the system both in the classical and quantum limit are very well reproduced by the resistively and capacitively shunted junction (RCSJ) model [3,4] (see Figure 1a). However, in some cases this simple model cannot be directly applied to JJs made from high critical temperature superconductors (HTS). The dynamics of HTS JJs are by far more complicated. The elaborate phenomenology is based on intrinsic properties, such as the unconventional *d*-wave order parameter and anisotropic charge transport [5].

Recent measurements performed on a $YBa_2Cu_3O_{7-x}$ (YBCO) JJ fabricated using a biepitaxial technique on a (110) $SrTiO_3$ (STO) substrate have shown that the classic and quantum dynamics of the phase difference across the JJ cannot be described by the conventional RCSJ model. Instead, an extended RCSJ model, including the effect of the stray capacitance and the stray inductance, has to be employed (see Figures 2a and 2b) [6, 7].

Moreover, in LTS quantum circuits large shunt capacitors and/or large inductors are intentionally used to influence the properties of the quantum circuit, e.g. large shunting capacitors are used to reduce the plasma frequency of phase qubits [8] or readout SQUIDs [9]. Other designs use an inductor and capacitor as an isolation network to protect the qubit from its low impedance environment causing dissipation [10, 11].

Our goal is a detailed study of those types of *LC* shunted junctions which are often found in real experiments. In this case, the dynamics of the junction are modified by the presence of the shell circuit and new effects have to be taken into account. In particular, two energy scales are introduced, which in specific cases can strongly differ from the plasma frequency of the isolated junction [7]. This has important implications for the quantum mechanical behavior of the system as a whole.

In this article, we systematically study the influence of an $L_S C_S$ shell circuit on the dynamics of a Josephson junction in the quantum limit. We fabricated Nb/Al-AlO$_x$/Nb Josephson junctions shunted by various on-chip inductances $L_S$ and on-chip capacitances $C_S$. The high yield of the Nb/Al-AlO$_x$/Nb technology allows fabricating circuits with extremely well defined junction parameters (critical current density and junction capacitance) as well as electromagnetic environment parameters $L_S$ and $C_S$, which at the moment is not possible with state of the art HTS JJ fabrication techniques.

The article is organized as follows:
In section II, we discuss the phase dynamics of a bare Josephson junction according to the RCSJ model. Thereafter, the effect of a shunting *LC* circuit on the phase dynamics of the whole system is described. Section III contains information about the fabrication of *LC* shunted Josephson junctions using Nb/Al-AlO$_x$/Nb technology. The *dc* electrical properties of the bare Josephson junction and the high frequency characterization of the bare *LC* resonator are presented in section IIIa and IIIb, respectively. The spectroscopic measurements of an *LC* shunted Josephson junction are summarized in section IV, followed by the conclusions in section V.

II. DYNAMICS OF AN *LC* SHUNTED JOSEPHSON JUNCTION: EXTENDED RCSJ MODEL

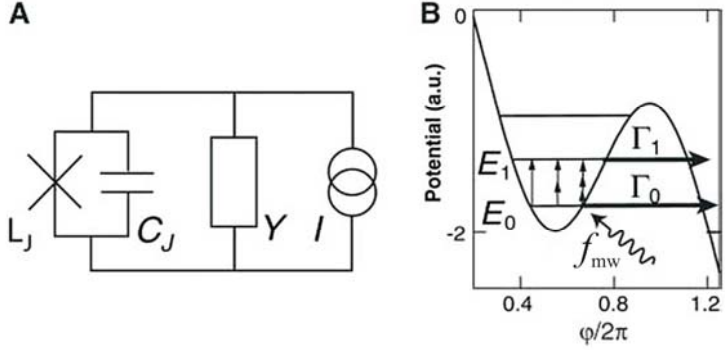

**Figure 1**: a) Circuit diagram of a current biased JJ in the RCSJ model. Damping of the JJ due to the environment and due to intrinsic effects is described by an admittance *Y*.
b) Energy levels in the potential of the current biased JJ. Microwave radiation induces a transition from the ground state to the first excited state having a larger escape rate than the ground state.

In this paragraph, we will first review aspects of the phase dynamics in a standard JJ, which can be described within the framework of the RCSJ model [12] (Figure 1a). Here, the junction inductance $L_J$ and capacitance $C_J$ act as an anharmonic *LC* resonator (at zero voltage) with resonance frequency $\omega_P = (L_J C_J)^{-1/2}$. Assuming a sinusoidal current phase relation, the Josephson inductance is given by $L_J = \phi_0/2\pi I_C \cos\varphi = L_{J0}/\cos\varphi$, where $I_C$ is the critical current, $\varphi$ is the gauge-invariant phase difference across the junction and $\phi_0 = h/2e$ is the superconducting flux quantum, with *e* being the elementary charge and *h* being Planck's constant. For a finite bias current, the fictitious phase particle with mass $m = (\phi_0/2\pi)^2 C_J$ may escape from a metastable well in the junction potential either by thermal activation or by tunneling through the potential barrier (see Figure 1b). This corresponds to the junction switching from the zero voltage state to a finite voltage state. At low temperature, the escape is dominated by tunneling [13]. For temperatures smaller than the energy level separation of the quasi-bound states in the well, only the ground state is populated. The quantum states can be observed spectroscopically by inducing a resonant transition between the ground state and the excited states by applying microwaves having frequencies $f_{0n} = (E_n - E_0)/h$ [14]. The bias current dependence of the transition frequency between the ground state and the first excited state can be approximated by the plasma frequency of small oscillations at the bottom of the well:

$$2\pi f_{01} \approx \omega_P = \sqrt{\frac{2\pi I_C}{\phi_0 C_J}\left(1-\left(\frac{I}{I_C}\right)^2\right)^{1/4}} = \omega_{P0}(1-\gamma^2)^{1/4}. \quad (1)$$

In the following, we will show how an $L_S C_S$ shell circuit connected to a JJ (see Figure 2a) affects the phase dynamics of the system as a whole.

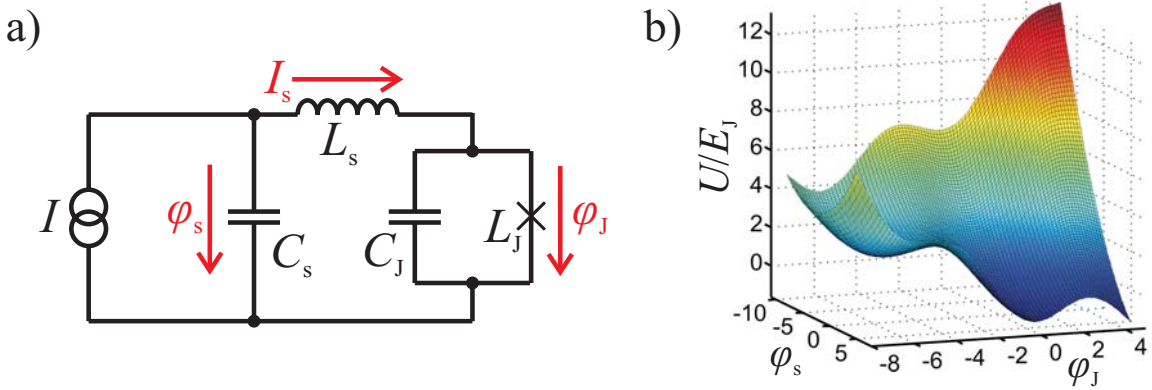

**Figure 2**: a) Circuit diagram of an *LC* shunted JJ. b) Two-dimensional potential $U$ (for details see [7]) of an *LC* shunted JJ for a bias current of $I = 0.5 I_C$, $C_S = 10 C_J$ and $L_S = 10 L_{J0}$. Here, $E_J = \Phi_0 I_C / 2\pi$ denotes the Josephson coupling energy.

In Figure 2a, the phase difference across the Josephson junction is given by $\varphi_J$. We can define $\varphi_S = (2\pi/\phi_0) I_S L_S + \varphi_J$, where $I_S$ is the current through the inductor. Neglecting dissipative elements in the circuit, we can write the normalized equation of motion as [7]

$$\ddot{\varphi}_J + \sin \varphi_J + \frac{\varphi_J - \varphi_S}{\beta} = 0, \qquad (2)$$

$$\chi^{-1} \ddot{\varphi}_S + \frac{\varphi_S - \varphi_J}{\beta} = \gamma. \qquad (3)$$

Here, time is normalized to the zero-bias plasma frequency of the Josephson junction $\omega_{P0} = \sqrt{2\pi I_C / \phi_0 C_J}$, $\gamma = I/I_C$ is the normalized bias current, $\beta = L_S / L_{J0}$ denotes the ratio between the shell inductance and the zero-bias Josephson inductance $L_{J0} = \phi_0 / 2\pi I_C$, and $\chi = C_J / C_S$ is the ratio between the junction capacitance and shell capacitance.

As can be seen from equations (2) and (3), the dynamics are described by two variables $\varphi_J$ and $\varphi_S$. Consequently, the simple picture of a fictitious phase particle moving in a one-dimensional potential, as shown in Figure 1a, is in general not valid anymore. Instead, the dynamics have to be described by a particle moving in a two-dimensional potential (see Figure 2b) [7], which results in two normal modes (or resonance frequencies) of the system. From equations (2) and (3), the resonance frequencies of the two normal modes of the *LC* shunted JJ (LCJJ) can be derived.

Here, we will only consider the limiting case $\chi \ll 1$, where the shell capacitance is much larger than the junction capacitance. This limiting case is common to many quantum devices containing Josephson junctions such as phase qubits [8] or readout SQUIDs of flux qubits [9] as well as HTS Josephson junctions [6].

The general expression for the upper and the lower resonant mode is given by [6]:

$$\omega_{\pm} = \omega_{P0} \sqrt{\frac{1 + \beta\sqrt{1-\gamma^2} + \chi \pm \sqrt{\left(1 + \beta\sqrt{1-\gamma^2} + \chi\right)^2 - 4\beta\chi\sqrt{1-\gamma^2}}}{2\beta}}.$$

For $\chi \ll 1$, the lower resonance mode can be approximated by

$$\omega_- = \frac{1}{\sqrt{(L_S + L_J)C_S}}, \qquad (4)$$

while the approximation for the upper resonance mode yields

$$\omega_+ = \sqrt{\left(\frac{1}{L_J} + \frac{1}{L_S}\right)\frac{1}{C_J}}. \qquad (5)$$

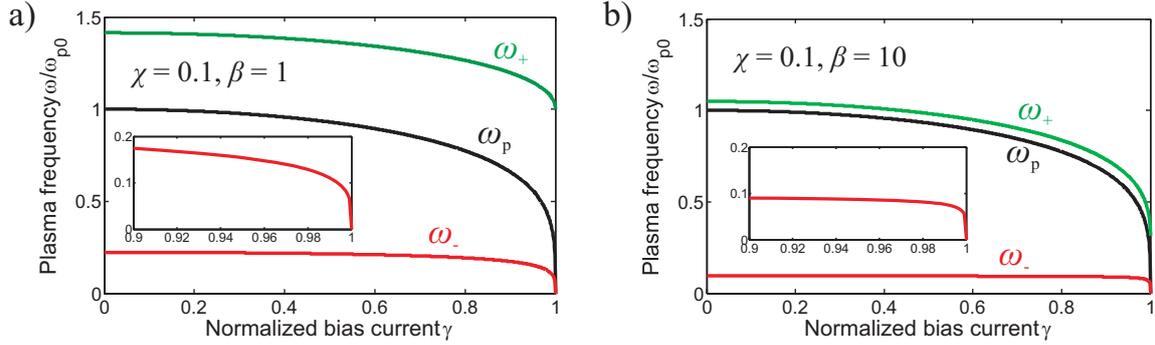

**Figure 3**: Plasma frequency $\omega_P$ for a single junction, upper normal mode $\omega_+$ of the LCJJ system and lower normal mode $\omega_-$ of the LCJJ system. The insets show a magnification of the lower mode close to $\gamma = 1$. a) For $\chi = 0.1$ and $\beta = 1$. b) For $\chi = 0.1$ and $\beta = 10$.

It follows directly that for $\beta \gg 1$, the upper resonant mode $\omega_+$ can be well approximated by the eigenfrequency of the Josephson junction $(L_J C_J)^{-1/2}$ whereas the lower resonance mode $\omega_-$ is given by the resonance frequency of the shell circuit $(L_S C_S)^{-1/2}$. This reflects the fact that for $\beta \gg 1$, the bare junction mode and the $LC$ mode are decoupled. From the point of view of novel phase dynamics and effects, the most interesting case is the one for $\chi \ll 1$ and not too large values of $\beta$. For $\beta = 1$ and $\beta = 10$, the theoretically expected eigenfrequencies $\omega_-$ and $\omega_+$ are shown in Figure 3 and compared to the pure Josephson plasma frequency $\omega_P$. The nonlinearity (non-vanishing bias current dependence) of both the upper and lower resonant mode clearly indicate that for $\beta \leq 10$, the bare junction mode and the $L_S C_S$ mode are strongly coupled.

It is worth pointing out that the lower resonant mode $\omega_-$ might be used for a phase qubit with improved dephasing times compared to a bare Josephson junction. This can be directly seen from the insets in Figures 3a and 3b. Dephasing is mainly caused by low frequency fluctuations in the biasing parameter, which in our case is the bias current. The dephasing rate is given by $\Gamma_\phi \approx \Delta I_n (\partial\omega/\partial I)$, where $\Delta I_n$ is the amplitude of the low frequency current noise. We see that a system having a flatter bias current dependence of the resonant mode will be less sensitive to bias current fluctuations. Therefore, a large value of $\beta$ is desirable. However, this value should not be too large, since for increasing $\beta$, the anharmonicity of the system is reduced, which makes a two-level approximation of the quantum system cumbersome.

## III. SAMPLE FABRICATION

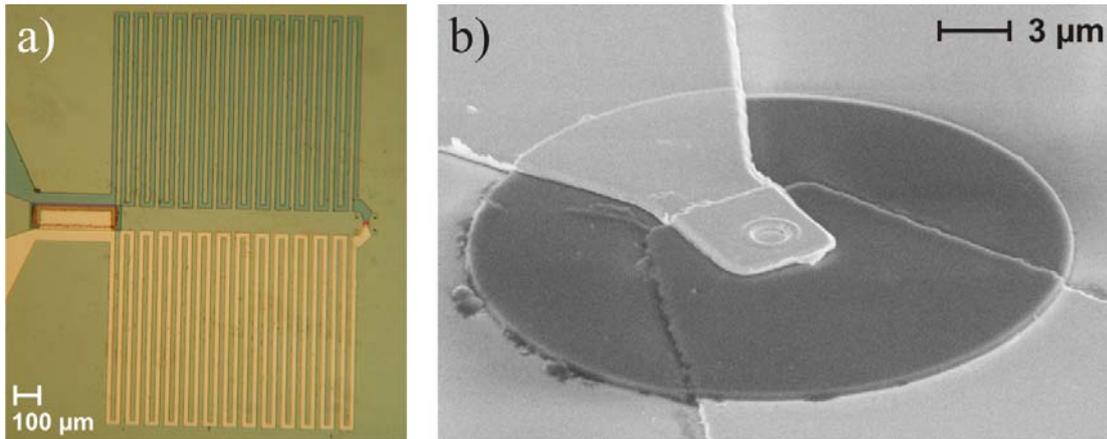

**Figure 4**: a) Photo of an LCJJ8 type sample. The capacitor is situated on the left while the JJ can be found on the right. The two meander lines connecting these elements act as inductors. The bottom electrode is shown on the upper part of the picture and has changed color due to the anodic oxidation. b) SEM image of the Josephson junction. The SiO insulation layer can be seen as the dark circle.

The samples were fabricated using a combined photolithography / electron-beam lithography process. First, the Nb/Al-AlO$_x$/Nb trilayer was DC magnetron sputtered *in-situ* with layer thicknesses of 100 nm / 7 nm / 100 nm. In order to obtain a critical current density of ~ 80 A/cm$^2$, the Al oxidation parameters were chosen accordingly. After that, the form of the bottom junction electrodes (including one half of the inductor and the bottom capacitor plate) were patterned by photolithography and reactive-ion-etching (RIE) as well as ion-beam etching. Then, the JJs were defined by a negative e-beam lithography process and subsequently etched by RIE. Using the same resist mask, the following anodic oxidation with a voltage of 20 V created an Nb$_2$O$_5$ layer of about 50 nm thickness for the first insulation layer of both the junction as well as the capacitor. The second insulating layer of SiO was patterned by a positive e-beam lithography process, thermal evaporation of the SiO onto the cooled samples and a subsequent lift-off process. Finally, the wiring layer (acting as the top electrode of the capacitor and the second part of the inductor) was deposited by DC magnetron sputtering and patterned by photolithography. A photo of a typical sample can be seen in Figure 4a, while an SEM image of the JJ itself is shown in Figure 4b. All junctions used within this work were round in shape and had a diameter of ~3 µm, meaning that they had a capacitance of $C_J$ ~0.4 pF and a Josephson inductance of $L_{J0}$ ~ 0.06 nH.

## IIIa) CHARACTERIZATION OF A SINGLE Nb/Al-AlO$_x$/Nb JOSEPHSON JUNCTION

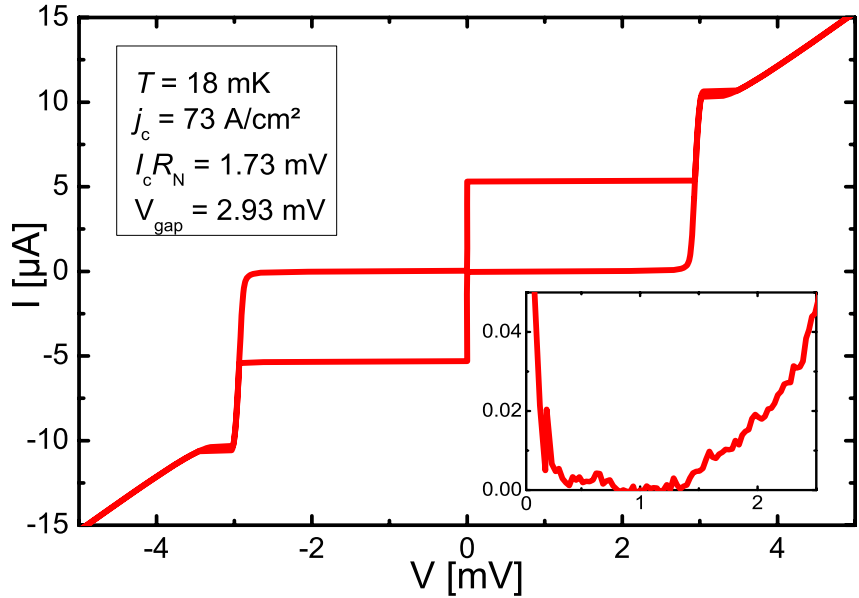

**Figure 5:** *IV* curve of a single junction having a diameter of $d = 3$ µm without the shell circuit. The high $I_cR_N$ and $V_{gap}$ values show that a very high quality was reached. The inset shows a magnication of the subgap branch as measured with a voltage bias configuration.

As a first step, single Nb/Al-AlO$_x$/Nb JJs with parameters identical to the ones used for the final spectroscopic measurements on *LC* shunted JJs were fabricated and characterized in order to see if a sufficient quality and reproducibility could be reached. As expected, the use of electron-beam lithography led to no significant spread in JJ sizes and critical currents $I_c$. The junctions were characterized at 4.2 Kelvin and the commonly used quality factor $R_{sg}/R_N$ was evaluated ($R_N$ being the junction's normal resistance and $R_{sg}$ its subgap resistance evaluated at a voltage of 2 mV). For all junctions, we found $R_{sg}/R_N > 35$, which indicates a very high quality. For further characterization at the temperature of the final quantum experiment, the junctions were cooled down to 18 mK. A typical current-voltage characteristic at this temperature is shown in figure 5. We found no excess currents and a high $I_cR_N$ product of 1.73 mV, meaning that we observe clear and clean Cooper pair tunneling [15]. Furthermore, measurements with a voltage bias setup showed very low leakage currents also at 18 mK, as can be seen in the inset of Figure 5. The high gap voltage of $V_{gap} = 2.93$ mV indicates that the Nb electrodes are of high quality. Furthermore, the modulation of the critical current with an external magnetic field (not shown) showed an Airy like pattern, as expected for a circular JJ. This indicates that the bias current is distributed homogeneously inside the junction.

## IIIb) CHARACTERIZATION OF THE $L_S C_S$ SHELL CIRCUIT

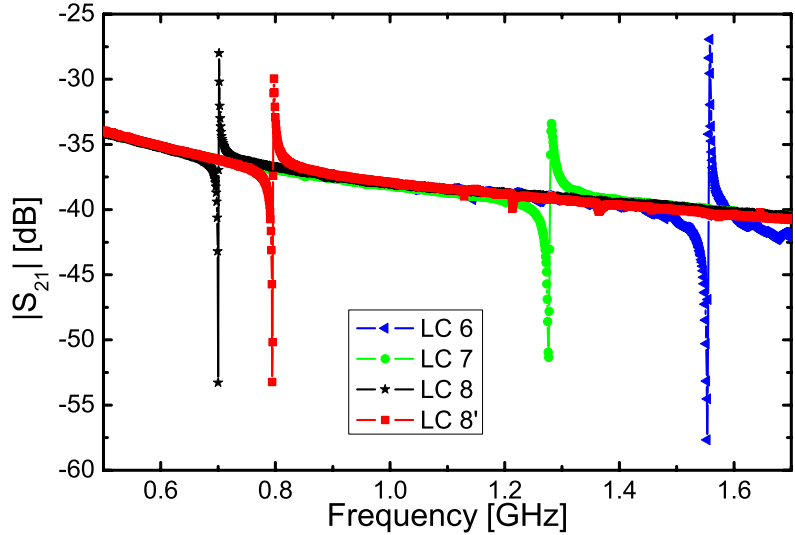

**Figure 6:** Measurements of shell circuits without the JJ. We found clear resonances at the expected frequencies with high loaded quality factors (see Table I).

In the following, the design, simulation and characterization of the external shell circuits will be described. For the calculation of the shell capacitance values, the simple plate capacitor formula was used. As we aimed for a ratio of $C_s/C_J = 10$, an overlay area of 0.02 mm$^2$ was chosen, so that $C_s = 3.7$ pF. The calculation of the inductance values, however, is more subtle. First, the kinetic inductance of the superconducting Nb lines has to be taken into account. Second, we found the inductance values to be frequency dependent for some of our desired shell circuits. We designed and simulated our structures with Sonnet [16] for a wide frequency range, taking Nb material parameters into account. The inductance value was taken at the point where the simulation frequency and the expected lower resonant mode frequency of the final LCJJ circuit matched. Here, we chose the lower resonant mode as it has a stronger deviation from the bare RCSJ plasma frequency for $\beta \gg 1$ (see Figure 3b). Eight different inductance geometries were designed, ranging from short straight lines to relatively narrow meanders (see Figure 4a), leading to $L_s$ values from 0.17 nH to 16 nH. This means that we could cover a wide range of $L_S/L_J$ ratios, systematically increasing from $\beta \sim 3$ for sample LCJJ1 to $\beta \sim 270$ for sample LCJJ8.

In order to see whether the simulation values were reliable, we fabricated *LC* shell circuits having the same design as the final samples, but replaced the Josephson junctions with a superconducting shortcut. Furthermore, a small gap was etched into one of the bias lines leading to the *LC* circuit, so that a coupling capacitor $C_c$ was formed, which would sufficiently decouple the *LC* circuit from the low-impedance environment. The experimental test of our designs was especially important for the structures with narrow meander lines, in order to exclude that a spurious capacitive coupling between the lines would shift the resonance or degrade its quality factor. For this purpose, we designed and fabricated a printed circuit board that allowed measurements up to about 2 GHz, which was enough to test the three smallest resonance frequencies and hence the three largest inductors. For one design, the resonance frequency was slightly varied by adjusting the SiO thickness, so that two samples

LC8 and LC8' were obtained. These preliminary measurements were carried out in liquid helium at 4.2 Kelvin. The chips were glued onto the printed circuit board and contacted by Al bonding wires. The PC board was contacted via mini-SMP connectors and 50 Ω cables to an Agilent Network Analyzer, which was used to measure the transmission *S*-parameter $|S_{21}|$. The results of the measurement are shown in Figure 6 and Table I. It can be seen that we got an excellent agreement between expected and experimentally determined resonance frequencies. This confirms the precision of our simulation method and excludes capacitive parasitics of any kind. We can conclude that if we have such excellent agreement even for the complex high-inductance structures (i.e. samples LC6 - LC8), the parameters of the low-impedance structures containing simple straight lines as inductors (i.e. samples LC1 - LC5), should also be well-defined. Furthermore, we obtained high loaded quality factors $Q_L$, which were probably only limited by the not well-defined $C_c$ values. This means that the shell circuits exhibit low damping, which is crucial for the observation of the underdamped dynamics of the LCJJ systems.

**Table I:** Measurement results for four investigated shell circuits without junctions. The experimentally determined resonance frequencies are within 3% of the ones calculated from the $L_s$ and $C_s$ design values. Additionally, high loaded quality factors $Q_L$ were obtained.

|       | $f_{res,design}$ [GHz] | $f_{res,meas}$ [GHz] | $Q_L$ |
|-------|------------------------|----------------------|-------|
| LC 6  | 1.60                   | 1.56                 | 777   |
| LC 7  | 1.27                   | 1.28                 | 138   |
| LC 8  | 0.72                   | 0.70                 | 506   |
| LC 8' | 0.80                   | 0.80                 | 236   |

IV. MEASUREMENT OF AN *LC* SHUNTED Nb/Al-AlO$_x$/Nb JOSEPHSON JUNCTION

The measurements of an LCJJ were performed in a dilution refrigerator with a base temperature of 15 mK. The *dc*-lines connecting the sample to the room temperature electronics were filtered with passive *RCL* filters at the 1K stage and with copper powder filters at the mixing chamber stage [17].
The design parameters of the investigated LCJJ system are given in Table II.

**Table II:** Design parameters of an *LC* shunted Josephson junction.

| sample | $j_C$ [A/cm$^2$] | $I_C$ [µA] | $d$ [µm] | $C_J$ [pF] | $C_S$ [pF] | $\chi$ [] | $L_{J0}$ [pH] | $L_S$ [nH] | $\beta$ [] |
|--------|------------------|------------|----------|------------|------------|-----------|---------------|------------|------------|
| LCJJ2  | 77               | 5.48       | 3.01     | 0.38       | 3.7        | 0.10      | 60            | 0.48       | 8.0        |

At $T \approx 20$ mK, which is clearly below the crossover temperature from the thermal to the quantum regime, microwaves were applied to the LCJJ in order to observe resonant activation of the phase dynamics [14]. These measurements allow analyzing the bias current dependence of the energy level separation.
For such spectroscopy experiments, it is likely that processes of higher order will also be observed. These can be multi-photon and/or multi-level processes [18]. This leads to transitions at microwave frequencies $2\pi f_{mw} \approx \frac{p}{q}\omega_{\pm}$. Other processes of higher order are mixed transitions like $2\pi f_{mw} \approx \frac{p}{q}\omega_{+} \pm \frac{m}{n}\omega_{-}$ ($p, q, m$ and $n$ are integers). The equality is only true for $p = 1$ and $m = 1$ as the anharmonicity of the potential causes the level spacing to shrink for increasing excitation level $p$ and $m$.

Spectroscopy measurements were carried out on sample LCJJ2 to reveal whether the two orthogonal resonant modes $\omega_{-}$ and $\omega_{+}$ would be observed instead of the plasma frequency of the single junction. The measurement results are shown in Figure 7. Here, we can clearly identify the contributions of the two resonance modes. In the same graph, theoretically calculated curves according to equations (4) and (5) are shown. It can be seen that the measurement data are in excellent agreement with the theoretical curves. Furthermore, no data points corresponding to the single junction were found, except where modes of the single junction and modes of the LCJJ system cross, which is visible in particular at half the plasma frequency $f_p/2$. Since no data points where found in regions of the spectrum where only single junction modes would be expected, we attribute the points at the crossings to the corresponding LCJJ modes. Altogether, our measurements provide strong evidence that the LCJJ system is very well described by theory and acts indeed as one single quantum system. This is remarkable since the circuit has a size of 200 × 650 μm$^2$.

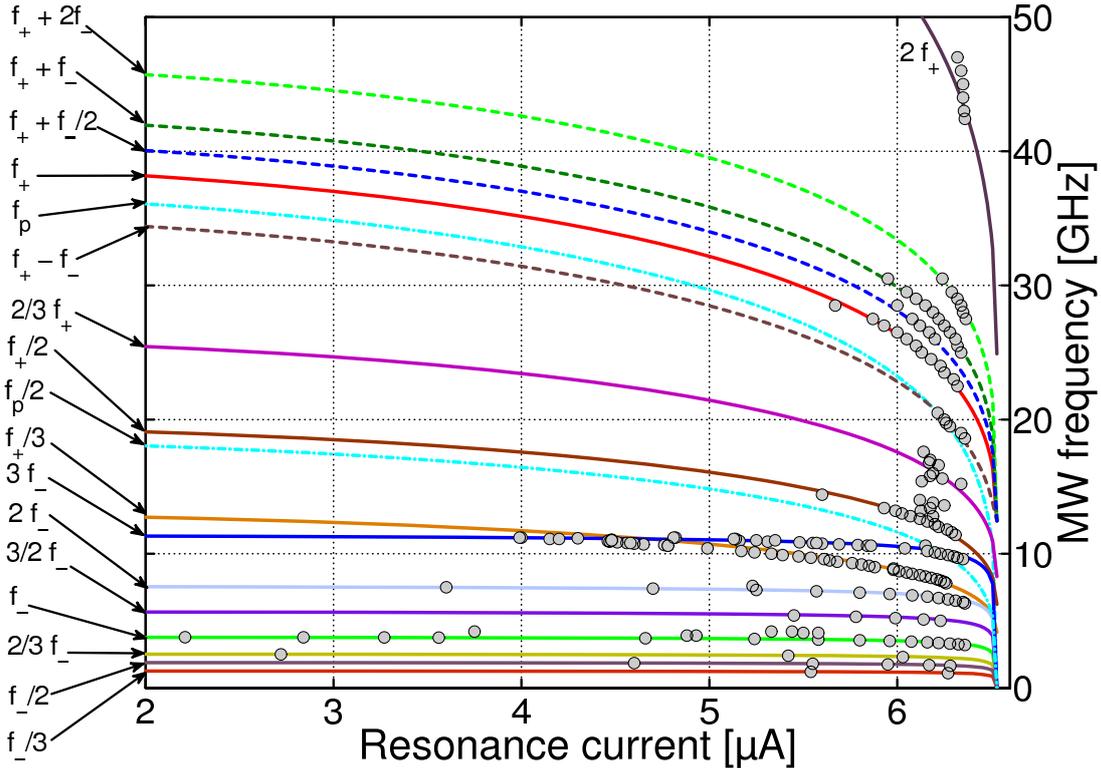

**Figure 7**: Spectroscopy data (gray points) of sample LCJJ2 without applied magnetic field. An excellent agreement between experiment on the one side and design values and theory on the other side is observed.

All theoretical curves in Figure 7 were calculated using one fixed set of parameters. For the junction these were a critical current of $I_C$ = 6.53 μA, and a Josephson capacitance of $C_J$ = 0.38 pF. For the shell circuit, the parameters accounted for $L_S$ = 0.43 nH and $C_S$ = 3.7 pF. That means that effectively, $\chi$ = 0.10 and $\beta$ = 8.5. The values used to fit the experimental spectroscopic data and the design values given in Table II are in excellent agreement. Only the critical current was slightly higher and the shell inductor slightly lower than planned. This shows that the parameters of our Josephson junction circuits are completely under control and the quantum properties of the circuits can be designed at will.

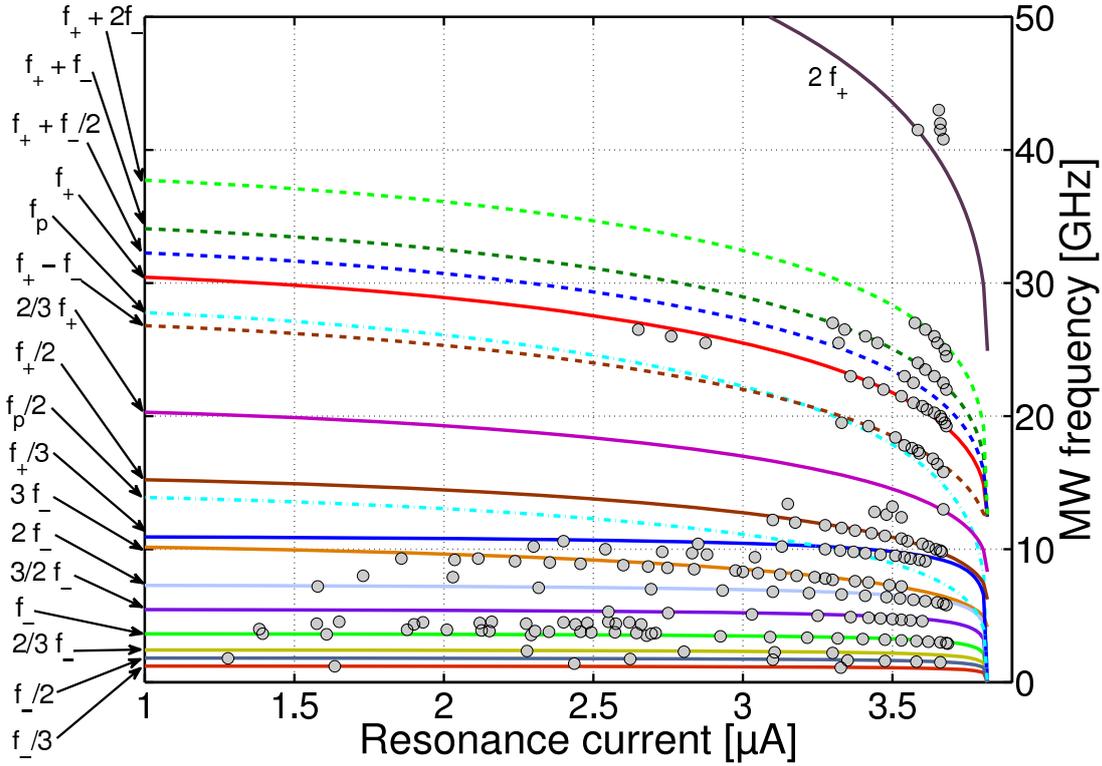

**Figure 8**: Spectroscopy data (gray points) of sample LCJJ2 with applied magnetic field, suppressing the critical current to $I_c = 3.82$ µA. The agreement between theory and experiment is excellent.

On the same LCJJ, the critical current was suppressed to $I_c = 3.82$ µA by applying a magnetic field and the same spectroscopy measurements were repeated. The measurement results are shown in Figure 8. All theoretical curves in this graph were calculated with the same parameters as in the absence of magnetic field. Again, we have an excellent agreement between measurement data and theoretical expectation. This further confirms our conclusion that the *LC* circuit strongly affects the phase dynamics of a Josephson junction.

V. CONCLUSIONS

We have shown that a Josephson junction shunted by an inductance and a capacitance acts indeed as one single quantum system. This is remarkable since the circuit has a size of $200 \times 650 \mu m^2$, which is larger than for most superconducting qubits. Furthermore, we found excellent agreement between the measurement data and the theoretically expected behavior calculated by using the design parameters. The findings show that such an LCJJ system has indeed two new energy scales instead of the energy scale of a single Josephson junction. Our results are important for the understanding of the dynamics in high-temperature Josephson junctions, where such shunting elements cannot be avoided when substrates with high dielectric constants are involved. Furthermore, the findings are important for superconducting quantum circuits such as quantum bits. When such systems are shunted capacitively by $C_s$, it is mostly assumed that $L_s = 0$ (which is definitively not the case since all electrical connections exhibit an inductance) and $\omega_P = \omega_- = (L_J C_S)^{-1/2}$. We have shown that $L_s$ needs to be considered resulting in $\omega_- = ([L_J + L_S]C_S)^{-1/2}$ and that there is a second energy scale in the system, namely $\omega_+ = \left(\left[1/L_J + 1/L_S\right]/C_J\right)^{1/2}$. This is an important result for quantum device design and operation.

VI. ACKNOWLEDGEMENTS

This work was partly supported by the DFG Center for Functional Nanostructures, project number B3.4, the Swedish Research Council (VR), EU STREP project MIDAS, the Knut and Alice Wallenberg Foundation. F.L. was supported by a grant from the Knut and Alice Wallenberg Foundation. Furthermore, we would like to thank M. Birk for help with inductor simulation, D. Bruch for design of the printed circuit board as well as J. Czolk and T. Wienhold for help with *LC* circuit characterization.

[1] T. van Duzer, *Principles of Superconductive Devices and Circuits*, 2nd ed. (Prentice Hall, Upper Saddle River, NJ, 1999).
[2] Y. Makhlin, G. Schön, and A. Shnirman, Rev. Mod. Phys. **73**, 357 (2001).
[3] W. C. Stewart, Appl. Phys. Lett. **12**, 277 (1968).
[4] D. E. McCumber, J. Appl. Phys. **39**, 3113 (1968).
[5] C.C. Tsuei, and J.R. Kirtley, Rev. Mod. Phys. **72**, 969 (2000).
[6] T. Bauch *et al.*, *Science* **311**, 57, (2006).
[7] G. Rotoli, T. Bauch, T. Lindström, D. Stornaiuolo, F. Tafuri, and F. Lombardi, Phys. Rev. B **75**, 144501 (2007).
[8] M. Steffen, M. Ansmann, R. McDermott, N. Katz, R. C. Bialczak, E. Lucero, M. Neeley, E. M. Weig, A. N. Cleland, and J. M. Martinis, Phys. Rev. Lett. **97**, 050502, (2006).
[9] A. Lupascu, C. J. M. Verwijs, R. N. Schouten, C. J. P. M. Harmans, and J. E. Mooij, Phys. Rev. Lett. **93**, 177006, (2004).
[10] A. J. Berkley, H. Xu, M. A. Gubrud, R. C. Ramos, J. R. Anderson, C. J. Lobb, and F. C. Wellstood, Phys. Rev. B **68**, 060502, (2003).
[11] J. Claudon, F. Balestro, F. K. J. Hekking, and O. Buisson, Phys. Rev. Lett. **93**, 187003, (2004).
[12] M. Tinkham, *Introduction to Superconductivity*, 2nd ed. (McGraw-Hill, New York, 2004).
[13] A. O. Caldeira and A. J. Leggett, Phys. Rev. Lett. **46**, 211 (1981).
[14] M. H. Devoret, D. Esteve, C. Urbina, J. Martinis, A. Creland, and J. Clarke, *in Quantum Tunneling in Condensed Media*, Y. Kagan and A. J. Leggett, (Eds. Amsterdam: North-Holland, 1992).
[15] V. Ambegaokar and A. Baratoff, Phys. Rev. Lett. **10**, 486 (1963)
[16] Sonnet Software Inc., 1020, Seventh North Street, Suite 210, Liverpool, NY 13088, USA.
[17] T. Bauch, F. Lombardi, F. Tafuri, A. Barone, G. Rotoli, P. Delsing, and T. Claeson, Phys. Rev. Lett. **94**, 087003 (2005).
[18] A. Wallraff, T. Duty, A. Lukashenko, A. V. Ustinov, Phys. Rev. Lett. 90, 037003 (2003).